# Elastic modeling and total energy calculations of the structural characteristics of 'free-standing', periodic, pseudomorphic GaN/AlN superlattices

Th. Karakostas[1], Ph. Komninou[1], and V. Pontikis[1,2,a]

(1) *School of Physics, Aristotle University of Thessaloniki, GR54124 Thessaloniki, Greece*

(2) *Training and Consulting on high-performance Materials (TCM), 92160 Antony, France*

**ABSTRACT**

The strain states of the components of pseudomorphic superlattices can be accurately modeled analytically through the application of linear elasticity. In this particular case of GaN/AlN 'free-standing' superlattices, the predictions derived from elastic modeling have been compared with total energy calculations of several systems made of components with varying thicknesses. We demonstrate that the elastic predictions for the lattice constants of the components align with the values obtained from their total energy counterparts, within the limits of computational errors and uncertainties. Furthermore, a phenomenological analysis of the elastic energy stored in the superlattices facilitates the evaluation of the excess energies associated with the interfaces present in these systems. The results mentioned above are briefly contrasted with findings reported in previous literature.

**I. INTRODUCTION**

The elements of pseudomorphic superlattices (SL), which consist of alternating thin layers of semiconducting materials, generally undergo strain as a result of the lattice mismatch between these constituents. Such nanostructures can be engineered to produce materials with customized electronic and optical characteristics that are distinct from their bulk equivalents by leveraging the strain-induced alterations of these properties. Specifically, strain can lower the effective mass of charge carriers, thereby influencing their mobility, and can either increase or decrease the bandgap or cause transitions between direct and indirect bandgaps. This method of deliberately applying strain to semiconducting materials, commonly known as strain engineering[1,2], is a component of the broader concept of gap engineering, which encompasses additional techniques aimed at controlling the bandgap of semiconductors, such as alloying and quantum confinement[3,4,5]. We extensively use total energy calculations to explore the

---

a) Author to whom correspondence should be addressed. Electronic mail: vassilis.pontikis@tcmat.org.

capabilities of short-period SL to provide materials with the desired electronic and optical properties, aiming at guiding the design of advanced optoelectronic devices[1-5].

In this study, we demonstrate that for any superlattice (SL), linear elasticity can accurately predict the strain states of both the SL and of its components without the need for computationally demanding total energy calculations, with the elastic moduli of the respective single-crystalline compounds serving as the only input. For illustration, GaN/AlN SLs are examined with components of varying thicknesses, and the elastic predictions of their dimensional characteristics are compared to the results of abinitio calculations. We observed that both methods produce identical results within the bounds of numerical errors and uncertainties. Moreover, the nearly perfect elastic behavior of the SL constituents enables a phenomenological analysis of the interfacial excess energies, providing quantitative insights into their energetic costs. Subsequently, Section II addresses the geometric characteristics of the systems under investigation and offers details regarding the total energy calculations that validate the elastic modeling. Section III introduces the elastic model and the resulting analytic expressions for the strain states and dimensional characteristics of the studied SL. It also includes the phenomenological analysis of interfacial excess energies, which arises from the integration of the elastic modeling with the outcomes of the total energy calculations. Finally, the results are discussed in Section IV, which also outlines the directions for ongoing work and includes some concluding remarks.

## II. MODEL AND COMPUTATIONS

### A. Model

Geometric models used as initial configurations for the calculations were prepared conformal to the experimental observations of superlattices and quantum wells, with reference axes x, y, and z spanning the crystallographic directions $[\bar{1}010]$, $[1\bar{2}10]$ and $[0001]$ of the wurtzite structure. The models consisted of a single unit cell in the basal plane (one atom) and m-GaN, n-AlN lattice cells along the [0001] c-direction.

### B. Total energy computations

We investigated energetic and structural properties of GaN/AlN superlattices via total energy calculations with the Quantum Espresso (QE) abinitio package[6,7,8] and the thermo_pw driver of routines[9] which compute abinitio material properties. The density functional of Wu et al.[10] was implemented using a Projector Augmented Wave (PAW) pseudopotential including a non-linear core correction[11]. This choice is justified by observing that it realistically reproduces experimental lattice constants and the elastic moduli of the single-crystalline compounds constitutive of the studied superlattices (Table I).

**TABLE I.** Computed properties of AlN, GaN single crystalline wurtzites: lattice parameters, a, c (nm), elastic moduli (GPa), and total energy, $E_{total}$ (eV). For the sake of the comparison, experimental data are also reported between parentheses[12,13].

| Quantity | *AlN* | *GaN* |
|---|---|---|
| a | 0.3114 (0.3113) | 0.3185 (0.3189) |
| c | 0.4987 (0.4982) | 0.5191 (0.5185) |
| $C_{11}$ | 382.1 (411±10) | 355.3 (390.0) |
| $C_{12}$ | 137.3 (149±10) | 126.0 (145.0) |
| $C_{13}$ | 107.0 (99.0±4) | 88.6 (106.0) |
| $C_{33}$ | 358.5 (389±10) | 394.2 (398.0) |
| $E_{total}$ | -1831.510 | -8302.308 |

The calculations were performed with the Broyden–Fletcher–Goldfarb–Shanno (BFGS) minimization algorithm[14] and cutoffs for the kinetic and potential energies set respectively equal at, $E_k$=140 Ry and $U_p$=1260 Ry. Moreover, the following convergence criteria have been adopted for the energy, $\delta E < 10^{-8}$ Ry, forces, $\delta F < 10^{-8}$ Ry/Bohr, and stresses, $\delta\sigma < 10^{-9}$ GPa. Confirmation that the above parameterization yields lattice constants and elastic moduli of the single crystalline binary compounds in good agreement with their experimental counterparts has been reported by previous work[15]. We obtained relaxed, strain-balanced (free-standing) bi-crystalline structures by minimizing the energy of initial configurations with full periodic boundary conditions (PBC) and vanishingly small global stress computed over the computational domain, $\bar{\bar{\Sigma}} \approx 0$. Values of the total energy and of the linear dimensions of the relaxed supercells are listed in Table II (Table A.1, Appendix A), including also previously reported values[15].

### III. ELASTIC MODEL

The main focus of this research is on SLs similar to those simulated via total energy calculations. It is essential to underscore that these have played a significant role in exploring how these nanostructures can be engineered to produce materials with customized electronic and optical properties that differ from those of their bulk alloy counterparts, which has been vital for the advancement of advanced optoelectronic devices. Particular attention has been directed here towards the so-called 'free-standing' pseudomorphic SLs constructed from m-GaN and n-Aln c-lattice cells within a parallelepiped computational domain, which is rendered pseudo-infinite through the implementation of PBC (§II.B), while ensuring that normal stress components remain vanishingly small (mechanical equilibrium condition).

The elastic modeling of such systems consists of applying Hooke's law (Appendix B) while ensuring that the total elastic energy stored in the SL is minimal, which would yield the strain states of the SL constituents and thereby their respective lattice constants. The mechanical equilibrium condition is expressed as follows:

$$\Sigma_1 = \Sigma_2 = \frac{(nc'\sigma'_1 + mc''\sigma''_1)}{(nc' + mc'')} = 0 \quad (1a)$$

$$\Sigma_3 = \sigma'_3 + \sigma''_3 = 0 \quad (1b)$$

where, the superscripts designate respectively AlN (') and GaN ("), with $\Sigma_1, \Sigma_2, \Sigma_3$ the normal components of the total stress acting on the SL, $\sigma'_1, \sigma''_1, \sigma'_3, \sigma''_3$ act on its constituents and c', c" the respective c-lattice parameters, whereas the usual short notation is used for the stress elements. Expressing $\sigma'_1, \sigma''_1, \sigma'_3, \sigma''_3$ via Hooke's law and substituting in Eqs. (1) yields:

$$nc'[(C'_{11} + C'_{12})e'_1 + C'_{13}e'_3] + mc''[(C''_{11} + C''_{12})e''_1 + C''_{13}e''_3] = 0 \quad (2a)$$

$$2C'_{13}e'_1 + C'_{33}e'_3 + 2C''_{13}e''_1 + C''_{33}e''_3 = 0 \quad (2b)$$

where, $e'_1 = e'_2, e'_3, e''_1 = e''_2, e''_3$, represent the principal deformations of AlN and GaN, $C'_{ij}, C''_{ij}$, the respective elastic moduli and c', c" the c-lattice homogenized parameters i.e. virtual parameters, ensure that the interfacial perturbations are homogeneously distributed over the corresponding spatial extensions along the c-axis of the two compounds. By substituting, $c' \approx c'_0(1 + e'_3)$ and $c'' \approx c''_0(1 + e''_3)$ into to eq. (2a) and by neglecting high-order terms, Eq. (2a) conveniently transforms into:

$$nc'_0[(C'_{11} + C'_{12})e'_1 + C'_{13}e'_3] + mc''_0[(C''_{11} + C''_{12})e''_1 + C''_{13}e''_3] \approx 0 \quad (2a')$$

where, $c'_0, c''_0$ are the c-lattice parameters of the reference single-crystalline compounds at the mechanical equilibrium. The stable equilibrium state of a given superlattice (n, m) corresponds to the minimum of the stored excess energy, $E_{el}$ (Appendix A):

$$\frac{\partial E_{el}}{\partial a_c} = 0 \quad (3)$$

with:

$$E_{el} \approx \frac{\sqrt{3}}{2} n a_c^2 \left\{ c'_0 \left[ (C'_{11} + C'_{12}){e'_1}^2 + \frac{1}{2}C'_{33}{e'_3}^2 + 2C'_{13}e'_1 e'_3 \right] + \frac{m}{n} c''_0 \left[ (C''_{11} + C''_{12}){e''_1}^2 + \frac{1}{2}C''_{33}{e''_3}^2 + 2C''_{13}e''_1 e''_3 \right] \right\} \quad (4)$$

whereas the deformation elements, $e'_1, e''_1$ relate with the common to the two compounds a-lattice parameter, $a_c$, (pseudomorphic relation) via:

$$e'_1 = \ln(a_c/a'_0), \quad e''_1 = \ln(a_c/a''_0) \quad (5)$$

where, $a'_0$, $a''_0$ are the a-lattice parameters of the single-crystalline compounds at the mechanical equilibrium.

In the presence of free-surface terminations along the common to the constituents' c-axis, the condition for mechanical equilibrium necessitates that $\sigma'_3 = \sigma''_3 = 0$, thereby ensuring that equation (1b) is inherently satisfied. When considering numerical simulations of superlattices with complete periodic boundary conditions (PBC) at the boundaries of the computational domain, the previously mentioned equation (1b) allows for a secondary solution, specifically $\sigma'_3 = -\sigma''_3$, which results in a deformation state that is significantly distinct from the previous scenario. In this instance, the elastic energy stored is greater than that observed in the first case, and this solution should be disregarded. From now on, the discussion pertains exclusively to the situation where $\sigma'_3 = \sigma''_3 = 0$. Ultimately, the current elastic modeling of superlattices is predicated on the assumption that the homogenization of c-lattice parameters throughout the thickness of the individual compounds is a valid consideration. The legitimacy of this assumption is demonstrated in section III, where total energy calculations for (n, m) AlN/GaN superlattices are presented.

While free-standing superlattices with free-surface terminations represent an intriguing field of study, these systems have not yet been experimentally developed or numerically modeled. Nevertheless, the aforementioned analysis indicates that the deformation states of computer-simulated superlattices under zero the normal components of the total stress, $\Sigma$, employing full periodic boundary conditions (PBC), mirror those of free-standing superlattices. This suggests that they may also serve as valid approximations of experimental systems, thereby supporting the subsequent elastic model.

In free-standing superlattices, eqs. (1b, 2b) hold due to the absence of stress components normal to the basal plane. Accordingly, the relevant 'Hooke's equation yields (Appendix B):

$$e'_3 = -2\frac{C'_{13}}{C'_{33}}e'_1, \quad e''_3 = -2\frac{C''_{13}}{C''_{33}}e''_1 \tag{6}$$

and by substitution in eq. (4) one obtains:

$$E_{el} \approx \frac{\sqrt{3}}{2}na_c^2\left[c'_0\left(C'_{11} + C'_{12} - 2\frac{{C'_{13}}^2}{C'_{33}}\right){e'_1}^2 + \frac{m}{n}c''_0\left(C''_{11} + C''_{12} - 2\frac{{C''_{13}}^2}{C''_{33}}\right){e''_1}^2\right] \tag{7}$$

On the other hand, eq. ('2a') leads to:

$$\beta = \frac{e'_1}{e''_1} = -\frac{m}{n}\frac{c''_0}{c'_0}\frac{C''_{11}+C''_{12}-2\frac{{C''_{13}}^2}{C''_{33}}}{C'_{11}+C'_{12}-2\frac{{C'_{13}}^2}{C'_{33}}} \tag{8}$$

By using the approximation:

$$e'_1 \approx \frac{a_c - a'_0}{a'_0}, \quad e''_1 \approx \frac{a_c - a''_0}{a''_0}, \tag{9}$$

the equilibrium a-lattice parameter common to the superlattice constituents is:

$$a_c = a'_0 \frac{\beta - 1}{\beta a'_0 / a''_0 - 1} \tag{10}$$

It is important to highlight that the analytical expression for β, derived from the minimization of elastic energy (as indicated in eqs. (3) and (7)), aligns perfectly with eq. (8) when quadratic deformation terms in the energy derivative are disregarded. The resulting expression for $a_c$ adheres to 'Hooke's law (eq. ('2a')) and effectively minimizes the elastic energy within the superlattice. Equation (8) indicates that β remains constant when the ratio m/n is maintained, suggesting that such free-standing superlattices will exhibit the same $a_c$ value at mechanical equilibrium. Additional structural parameters can be readily determined: $c' = c'_0 exp(e'_3)$, $c'' = c''_0 exp(e''_3)$, $L_{SL} = nc' + mc''$, with the latter denoting the thickness of the computational box along the c-direction [0001]. Lastly, it is noteworthy that the elastic energy stored (as expressed in (eq. (7)) is a linear function of n provided that the ratio m/n remains constant.

## IV. TOTAL ENERGY CALCULATIONS

The elastic models examined in Section II are validated using precise structural and elastic energy data derived from total energy calculations involving various GaN/AlN superlattice models. This validation approach is unaffected by uncertainties that might arise from the use of experimental data, while the extensive range of superlattice models available for computer simulations is typically unattainable through experimental means. The subsequent sections will elaborate on the models and computations employed, the methodology for extracting elastic energy values from the total energy of relaxed superlattices, and the outcomes of the comparison between the elastic models and the simulated superlattices.

### A. Strain and stress states of simulated superlattices

The validation of the elastic modeling necessitates the confirmation that its predictions regarding the internal strain and stress states of the various superlattices under investigation align with the values derived from their computational counterparts through total energy calculations. Specifically, this section focuses on the deformation states and lattice constants of the relaxed components, while the stored elastic energy in the superlattice will be examined in detail in the subsequent section.

The a and c-lattice parameters of the relaxed superlattices are used to determine the normal components of strains. The first is a direct outcome of the total energy calculations, whereas the latter is readily provided by exploring interatomic distances between nitrogen atomic planes along the [0001] crystallographic direction. Figure 1 displays such a profile for the system (m,n) = (6,6) taken as a generic example of the procedure yielding the c-lattice parameter values in the bulk of the SL components. We observe that interfacial relaxations of interplanar distances are practically negligible in AlN whereas in GaN these extend over a single interplanar distance. The predictions of the elastic model are also reported in this figure (dashed lines), $c' = c'_0 exp(e'_3)$, $c'' = c''_0 exp(e''_3)$ (eqs. 6, 12, Appendix A, A.2) to demonstrate that these are in fair to good agreement with the QE values. The corresponding strain elements, $e'_3$, $e''_3$ are plotted in figure 2 as functions of the common a-lattice constant and are in fair agreement with the predictions of the elastic model. Figure 3 presents the length of the computational box containing the studied superlattices along the c-direction as a function of the length of the corresponding unstrained mechanical mixture i.e. without interfacial perturbations. It shows that the elastic prediction (full line and open triangles) aligns fairly well with the QE calculations (open circles), consistent with the satisfactory prediction of $c'$, $c''$, and that the thickness of the corresponding unstrained mechanical mixture conveniently approximates LSL. Similarly, the elastic prediction of the a-lattice parameter (eq. (10)) is in excellent agreement with the QE calculations, as shown by Fig. 4.

A plane stress condition arises in the relaxed superlattices as a result of the elimination of the stress components $\sigma'_3$, $\sigma''_3$, which ensures their mechanical stability. Figures 5a and 5b illustrate the relationship between the tangential stress elements, $\sigma'_1 = \sigma'_2$, $\sigma''_1 = \sigma''_2$ along with the associated hydrostatic pressures, as functions of the ratio m/n. It is found that the elastic model predictions align perfectly with the results obtained using 'Hooke's law from the QE calculations of $e'_1 = e'_2$, $e'_3$, $e''_1 = e''_2$, $e''_3$. It can be deduced from equation (7) that the assessments performed, using both approaches, for the elastic energy stored in the SL yield results that are closely aligned, within numerical errors, as observed. This suggests that the representations of its elastic state through elastic modeling or total energy calculations are equivalent, despite the fact that the homogenization of interfacial perturbations is relevant in the former.

The present elastic modeling offers a reasonably accurate prediction of the strain and stress states in relaxed SLs, and the approximation that involves homogenizing the deformations along the c-direction effectively represents interfacial relaxations.

## B. Elastic and interfacial energies

The total energy of the simulated superlattices, denoted as $E_{total}$, can be expressed as follows:

$$E_{total} = E_{mm}(m,n) + E^{el}_{bulk}(m,n) + E_{int}(m,n) + E^{el}_{int}(m,n) \quad (11)$$

In this equation, $E_{mm}(m,n) = mE_{GaN} + nE_{AlN}$ is the energy of the mechanical mixture, where $E_{GaN}$, $E_{AlN}$ represent the energies of the SL constituents in their fundamental states (refer to Table I). Furthermore, $E_{int}(m,n)$ represents the interfacial excess energy of chemical origin, while $E^{el}_{bulk}(m,n)$ denotes the elastic energy stored within the bulk nitrides, assuming they are uniformly deformed to satisfy the pseudomorphic condition that requires both to adopt the same a-lattice constant. It is expected that the former will exhibit minimal variation with changes in the common a-lattice constant ac, or, correspondingly, the ratio m/n (refer to eqs. (8) and 10), since the resulting alterations in bond lengths at the interfaces are negligible. Lastly, $E_{int}^{el}(m,n)$ quantifies the anticipated deviation of the interfacial regions from the homogeneous elastic energy of the bulk, $E_{bulk}^{el}(m,n)$, which aligns with the distinctly differentiated profiles of interplanar distances along the c-direction (see Fig. 1). It is important to underline that the interfacial energy's unknown dependence on m and n reflects the reality that, at small thickness limits, interfacial perturbations overlap[15]. Moreover, for thicker SL constituents, their effective extension cannot be quantitatively extracted from the profiles of interfacial distances along the c-axis (Fig. 1).

Given that $E_{mm}(m,n)$ is the dominant factor among the unknown terms in equation (11), $E_{int}(m,n)$ and $E_{int}^{el}(m,n)$, it is anticipated that $E_{total} = E_{total}(m,n)$ will exhibit a linear variation with respect to n, while the ratio m/n remains constant. This claim holds true within the bounds of numerical uncertainties, as demonstrated by figure (6a). Conversely, given that interfacial relaxations demonstrate independence from the thicknesses of the constituents for values m>2 and n>2[15], the total excess energy, $\delta E(m,n)$, should also exhibit a linear variation with n whenever the ratio m/n remains constant:

$$\delta E(m,n) = E_{total} - E_{mm}(m,n) - E_{bulk}^{el}(a_c,m,n) \qquad (12)$$

Figure (6b) proves that this prediction holds true. Combining eqs. (11) and (12) one obtains:

$$\delta E(m,n) = E_{int}(m,n) + E_{int}^{el}(m,n) \qquad (13)$$

The linear least-squares fits shown in Fig. 6b provide the values of $E_{int}(m,n)$ and $E_{int}^{el}(m,n)/n$, which are detailed in Table II. From this table, it is evident that, as anticipated, the $E_{int}(m,n)$ values are nearly identical, also indicating m/n independence with minor fluctuations, the source of which cannot be definitively determined. The corresponding total interfacial energy values for the examined superlattices are presented in the final column, resulting in an average $\bar{\gamma} = 25.06$ mJ/m², a reasonable figure that aligns with the established stability against decohesion of such superlattices and with prior calculations[15].

**Table II.** The parameters $E_{int}(m,n)$ and $E_{int}^{el}(m,n)/n$, obtained via linear least-squares fittings of the excess energy $\delta E(m,n)$ as a function of n (refer to Fig. 6b), are provided for specific values of the ratio m/n. The associated common a-lattice parameter values, denoted as <$a_c$>, represent the averages of the fluctuating values listed in Table III (Appendix B). The last column provides the values of the total interfacial energy contained in the studied systems.

| m/n | $E_{int}(m,n)$ (eV) | $E_{int}^{el}(m,n)$ (eV) | <$a_c$> (nm) | $\gamma$ (mJ/m$^2$) |
|---|---|---|---|---|
| 0.2 | 0.01167 | 4.939 10$^{-3}$ | 0.3126 | 22.09 |
| 0.25 | 0.01300 | 5.539 10$^{-3}$ | 0.3128 | 24.58 |
| 1/3 | 0.01367 | 6.536 10$^{-3}$ | 0.3132 | 25.78 |
| 0.5 | 0.01335 | 8.523 10$^{-3}$ | 0.3138 | 25.15 |
| 1.0 | 0.01513 | 1.303 10$^{-2}$ | 0.3150 | 28.21 |
| 2.0 | 0.01375 | 2.041 10$^{-2}$ | 0.3162 | 25.44 |
| 3.0 | 0.01353 | 2.626 10$^{-2}$ | 0.3167 | 24.96 |
| 5.0 | 0.01321 | 3.769 10$^{-2}$ | 0.3173 | 24.27 |

## IV. DISCUSSION AND CONCLUSIONS

This research illustrates that linear elasticity, when utilized with single-crystal elastic moduli, offers a strong and computationally efficient framework for forecasting the structural properties of free-standing, pseudomorphic GaN/AlN superlattices. Analytical predictions regarding strain partitioning and relaxed lattice parameters are corroborated by total energy calculations from Quantum ESPRESSO, showing consistency within numerical uncertainty over a wide spectrum of layer thickness ratios.

The overall excess energy of the superlattice (SL) is defined as the aggregation of energetic components that correspond to the mechanical mixture, the chemical excess energy at the interface, the elastic energy retained within the bulk of the SL materials, and the elastic energy accumulated in the interfacial areas beyond their bulk equivalents. This methodology produces the total interfacial energies for the analyzed SLs, aligning with previous studies[15] and thereby supporting the phenomenological breakdown. Furthermore, comparing the analytical predictions of the stress states in the SLs with their total energy equivalents confirms the validity of the homogenization approximation of interplanar distances along the c-axis that underpins the proposed elastic model. In addition to its predictive precision, this model provides a practical benefit for material design by facilitating the swift estimation of strain states and stored elastic energies, eliminating the necessity for extensive density functional theory (DFT) computations, which is advantageous for high-throughput screening of SL configurations. Its ability to yield trustworthy interfacial excess energies, when integrated with total energy computations, also provides insights into interface stability. This understanding is crucial for optimizing growth conditions and minimizing defect formation.

Future research aims to expand the model by incorporating temperature effects, piezoelectric coupling, and polarity, while also investigating its relevance to alloyed or disordered systems. The integration with electronic structure calculations may significantly improve its usefulness in device-oriented design. In summary, the synergy of linear elasticity and targeted first-principles calculations provides a robust and scalable method for comprehending and manipulating the structural energetics of III-nitride superlattices.

In conclusion, this document presents a thorough examination of the crossroads of computational materials science and semiconductor physics. It illustrates that a precisely constructed linear elastic model, refined and confirmed through targeted first-principles calculations, serves as a robust predictive instrument that provides a scalable, efficient, and physically insightful approach for the rational design of next-generation strained semiconductor nanostructures aimed at advanced optoelectronic and quantum technologies.

## ACKNOWLEDGMENTS

We thank the High-Performance Computing Infrastructure and Resources Center of the Aristotle University of Thessaloniki for the generous supply of computing resources.

This research was performed while V. Pontikis was visiting the School of Physics, Aristotle University of Thessaloniki, Thessaloniki, GR54124, Greece.

## AUTHOR DECLARATIONS

**Conflict of Interest**

The authors have no conflicts to disclose.

**Author Contributions**

**Theodoros Karakostas:** Conceptualization (equal); Data curation (equal); Formal analysis (equal); Investigation (lead); Validation (lead); Review & editing (equal). **Philomela Komninou**: Conceptualization (equal); Formal analysis (equal); Investigation (equal); Validation (equal); Review & editing (equal). **Vasillis Pontikis:** Conceptualization (equal); Data curation (equal); Formal analysis (lead); Methodology (lead); Investigation (lead); Validation (equal); Writing – review & editing (lead);

# APPENDIX A: Calculated energetic and structural data of GaN/AlN superlattices

**Table A.1** Total energy $E_c$ (eV) and linear dimensions $a_c$, $L_{SL}$ (nm) of the studied, relaxed superlattices.

| m | n | $E_c$ (eV) | $a_c$ (nm) | $L_{SL}$ (nm) |
|---|---|---|---|---|
| 0.5 | 0.5 | -5066.875 | 0.3148 | 0.5101 |
| 0.5 | 1.0 | -5982.622 | 0.3137 | 0.7595 |
| 0.5 | 1.5 | -6898.373 | 0.3131 | 1.0089 |
| 0.5 | 2.0 | -7814.126 | 0.3128 | 1.2573 |
| 0.5 | 2.5 | -8729.880 | 0.3126 | 1.5076 |
| 0.5 | 3.0 | -9645.632 | 0.3124 | 1.7569 |
| 0.5 | 3.5 | -10561.39 | 0.3123 | 2.0062 |
| 0.5 | 4.0 | -11477.14 | 0.3122 | 2.2555 |
| 0.5 | 4.5 | -12392.89 | 0.3121 | 2.5049 |
| 1.0 | 0.5 | -9218.019 | 0.3161 | 0.7697 |
| 1.0 | 1.0 | -10133.76 | 0.3149 | 1.0192 |
| 1.0 | 2.0 | -11965.26 | 0.3138 | 1.5180 |
| 1.0 | 3.0 | -13796.76 | 0.3132 | 2.0166 |
| 1.0 | 4.0 | -15628.26 | 0.3128 | 2.5153 |
| 1.0 | 5.0 | -17459.77 | 0.3126 | 3.0139 |
| 1.5 | 0.5 | -13369.17 | 0.3167 | 1.0292 |
| 2.0 | 0.5 | -17520.32 | 0.3171 | 1.2888 |
| 2.0 | 1.0 | -18436.05 | 0.3161 | 1.5384 |
| 2.0 | 2.0 | -20267.53 | 0.3150 | 2.0372 |
| 2.0 | 4.0 | -23930.53 | 0.3136 | 3.0346 |
| 2.0 | 6.0 | -27593.53 | 0.3132 | 4.0320 |
| 2.0 | 8.0 | -31256.54 | 0.3129 | 5.0292 |
| 2.0 | 10.0 | -34919.55 | 0.3126 | 6.0265 |
| 2.5 | 0.5 | -21671.47 | 0.3173 | 1.5483 |
| 3.0 | 0.5 | -25822.62 | 0.3175 | 1.8078 |
| 3.0 | 1.0 | -26738.35 | 0.3167 | 2.0575 |
| 3.0 | 3.0 | -30401.30 | 0.3150 | 3.0551 |
| 3.0 | 6.0 | -35895.80 | 0.3138 | 4.5512 |
| 3.0 | 8.0 | 0.000000 | 0.3134 | 5.5485 |
| 4.0 | 0.5 | -34124.92 | 0.3177 | 2.3269 |
| 4.0 | 2.0 | -36872.11 | 0.3162 | 3.0755 |
| 4.0 | 4.0 | -40535.08 | 0.3150 | 4.0731 |
| 4.0 | 7.0 | 0.000000 | 0.3134 | 5.5485 |
| 4.0 | 8.0 | -47861.07 | 0.3138 | 6.0678 |
| 4.5 | 0.5 | -38276.19 | 0.3178 | 2.5868 |
| 5.0 | 1.0 | -43342.95 | 0.3173 | 3.0956 |
| 5.0 | 2.0 | 0.000000 | 0.3167 | 3.5904 |
| 5.0 | 5.0 | -50668.86 | 0.3150 | 5.0910 |
| 5.0 | 7.0 | -54331.84 | 0.3144 | 6.0884 |
| 5.0 | 10.0 | -59826.34 | 0.3138 | 7.5844 |
| 6.0 | 2.0 | -53476.71 | 0.3168 | 4.1137 |
| 6.0 | 3.0 | -55308.18 | 0.3162 | 4.6125 |
| 6.0 | 6.0 | -60802.63 | 0.3150 | 6.1089 |
| 7.0 | 5.0 | -67273.43 | 0.3156 | 6.1275 |
| 7.0 | 7.0 | -70936.40 | 0.3150 | 7.1269 |

## APPENDIX B: 'Hooke's equations

In the wurtzite crystalline structure and with reference to the crystallographic axes [$\bar{1}010$], [$1\bar{2}10$] and [0001], 'Hooke's equations relate principal elements of stress $\sigma_i$ and principal deformation $e_i$ (in short notation, i=1-6) as follows:

$$\sigma_1 = \sigma_2 = (C_{11} + C_{12})e_1 + C_{13}e_3 \tag{A.1}$$

$$\sigma_3 = 2C_{13}e_1 + C_{33}e_3 \tag{A.2}$$

where $C_{ij}$ are the elastic moduli. With vanishing the remaining deformation elements, $e_4 = e_5 = e_6 = 0$, the associated elastic energy is given by:

$$W = (C_{11} + C_{12})e_1^2 + \frac{1}{2}C_{33}e_{33}^2 + 2C_{13}e_1e_3 \tag{A.3}$$

[13] M. Leszcynski, H. Teisseyre, T. Suski, I. Grzegory, M. Bockowski, J. Jun and S. Porowski, "Lattice parameters of gallium nitride", Appl. Phys. Lett. **69** 73-75 (1996).

[14] R. Fletcher, *Practical Methods of Optimization*, 2nd ed., ISBN 978-0-471-91547-8 (John Wiley & Sons, New York, USA, 1987).

[15] Th. Karakostas, Ph. Komninou and V. Pontikis, "Energetics of interfaces ans strain partition in GaN/AlN pseudomorphic superlattices", Crystals **13**, 1272 (2023).

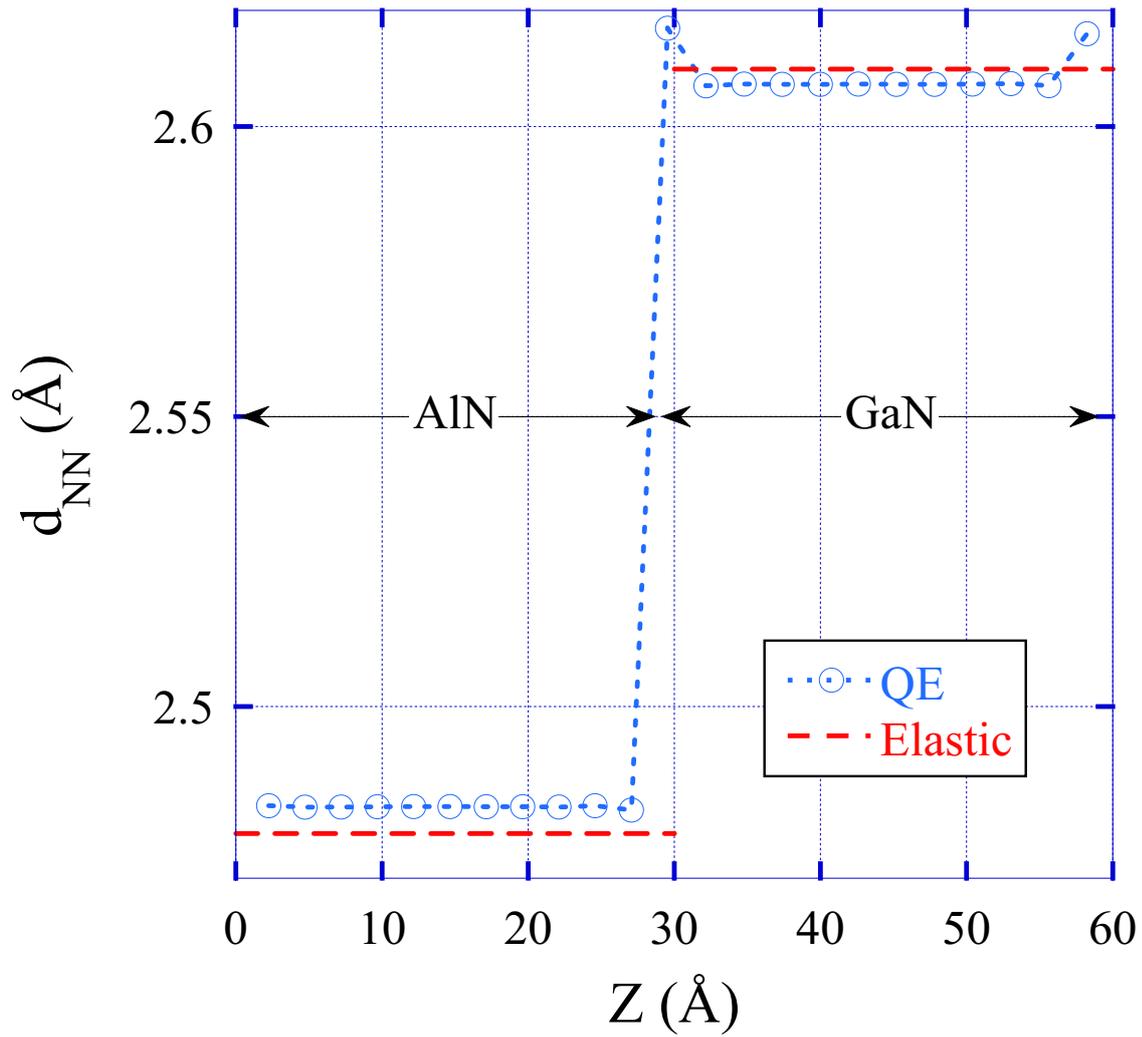

**Figure 1** – Interplanar distance $d_{NN}$ between consecutive nitrogen (0001) planes in a SL (m,n)=(6,6). In AlN, interfacial perturbations are minimal, while in GaN, these perturbations extend over one interplanar distance at the interfaces. The total energy calculations from QE (open circles) and the predictions from the elastic model (dashed lines) are presented. The dotted line serves as a visual aid.

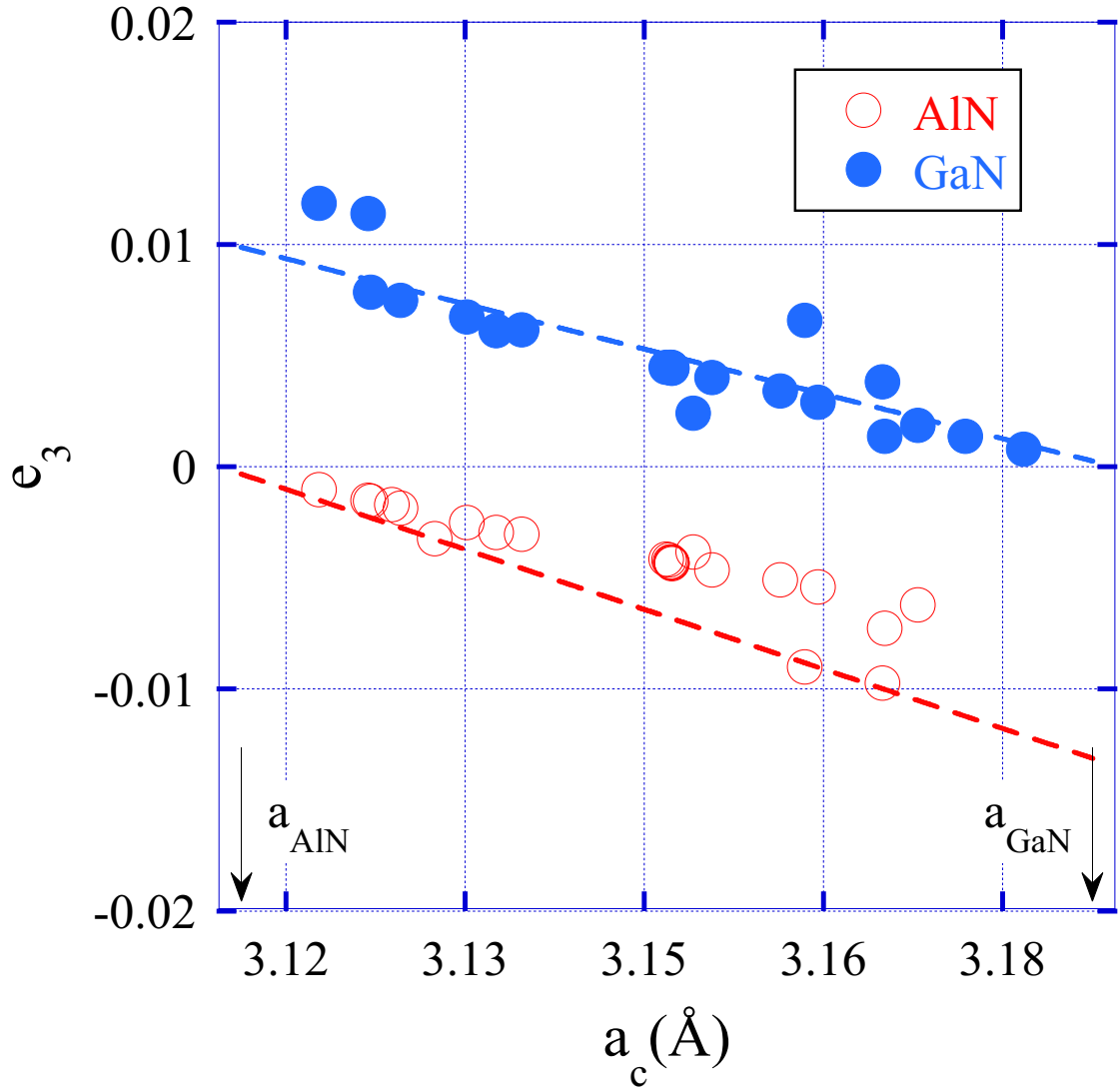

**Figure 2** – Normal deformation component $e_3$ of interplanar distances along the (0001) axis of the SL as a function of $a_c$. The elastic model is represented by dashed lines, while the total energy calculations are indicated by open circles for AlN and filled circles for GaN. Vertical arrows denote the values of $a_c$ for the single crystalline compounds at their fundamental states.

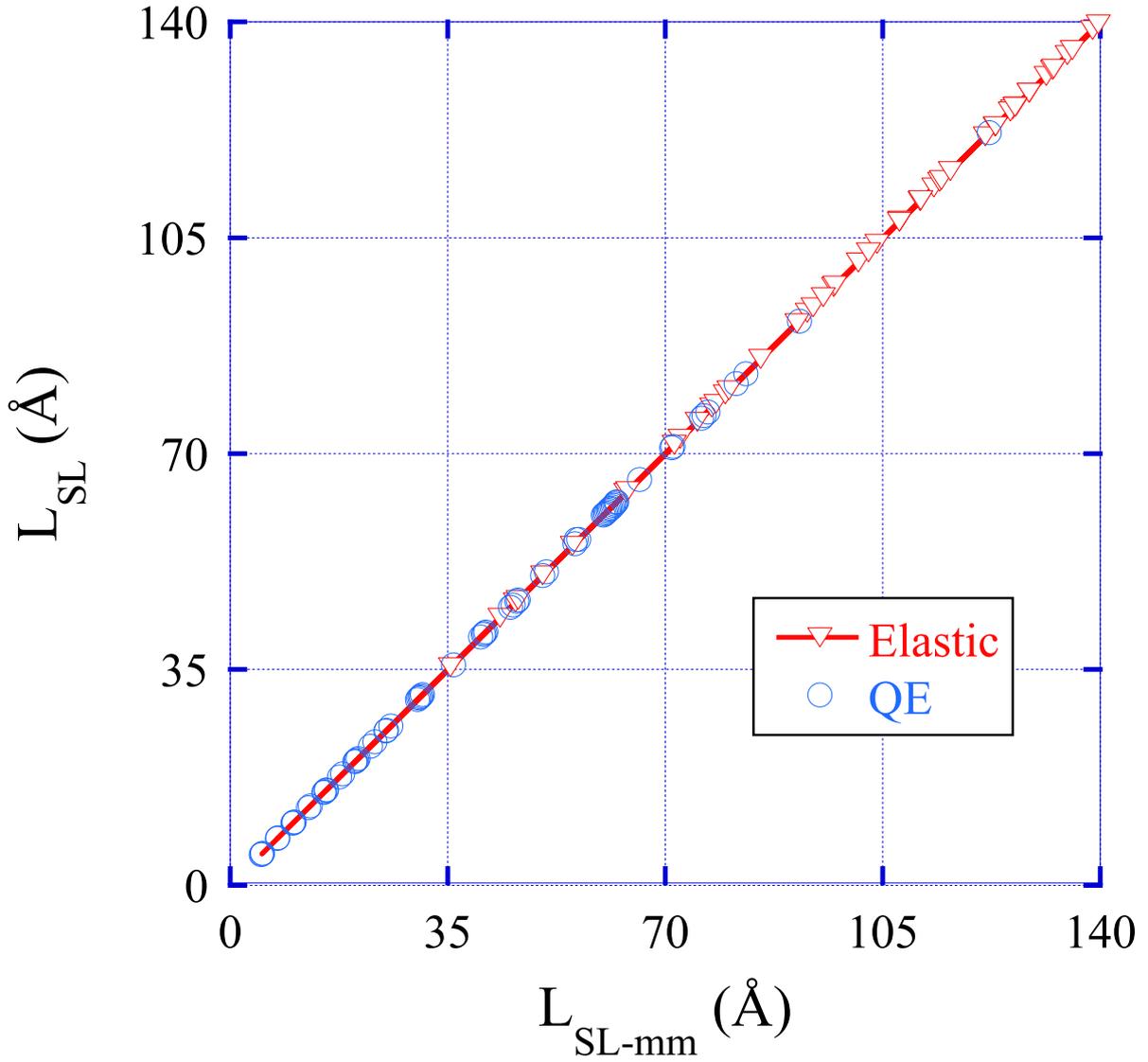

**Figure 3** – The thickness of different simulated superlattices (m,n) is presented as a function of the thickness of the associated mechanical mixture. The solid line and open triangles represent the elastic model, while the open circles denote total energy calculations (QE).

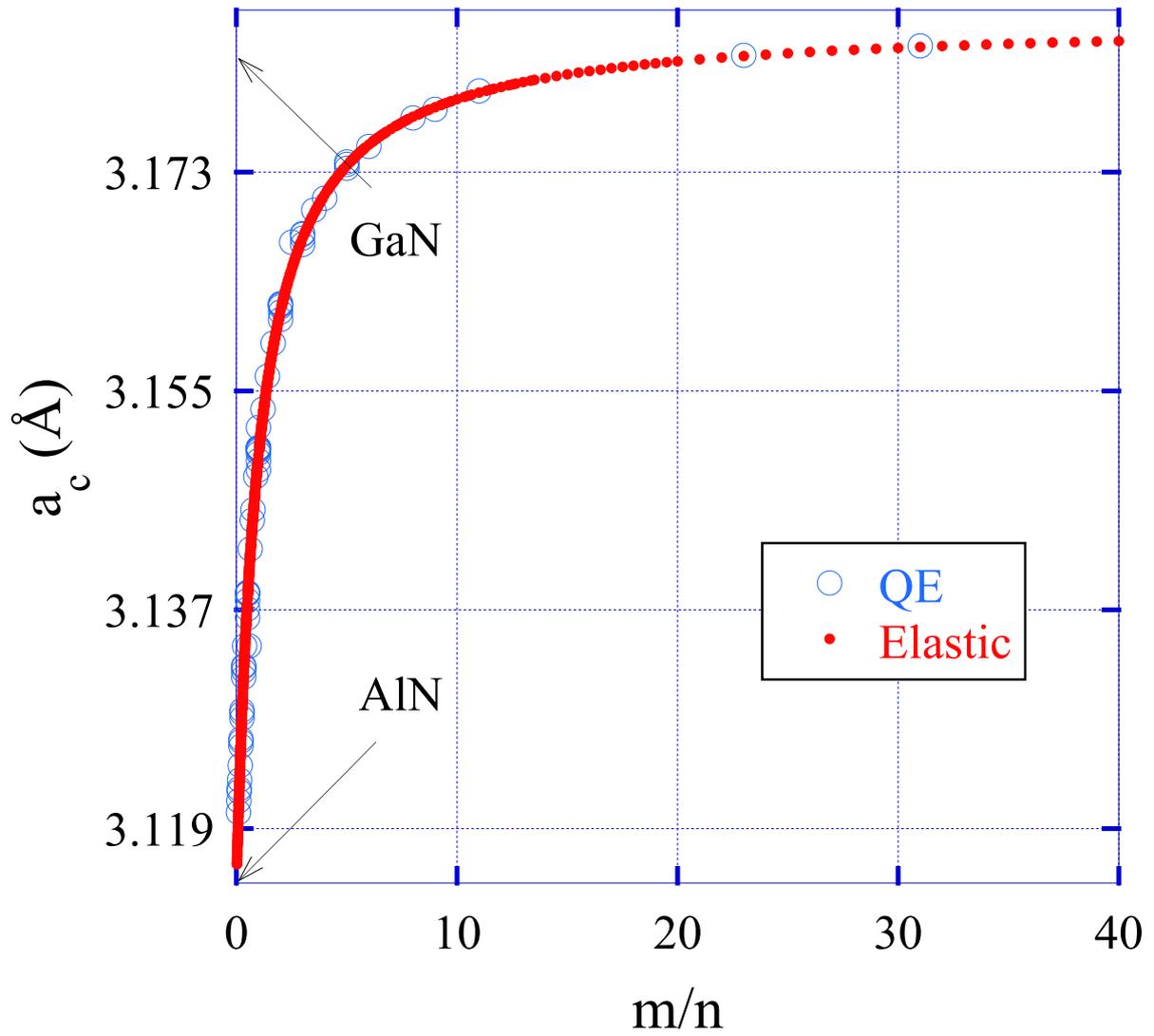

**Figure 4** – Common lattice parameter $a_c$ of the two nitrides as a function of the ratio m/n. Solid dots represent the elastic model, while open circles denote the total energy calculations (QE). The arrows indicate the a-lattice parameters of the two single crystalline nitrides in their fundamental states.

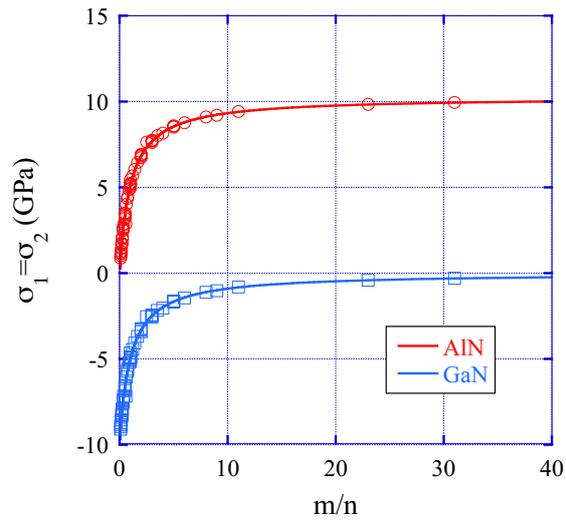 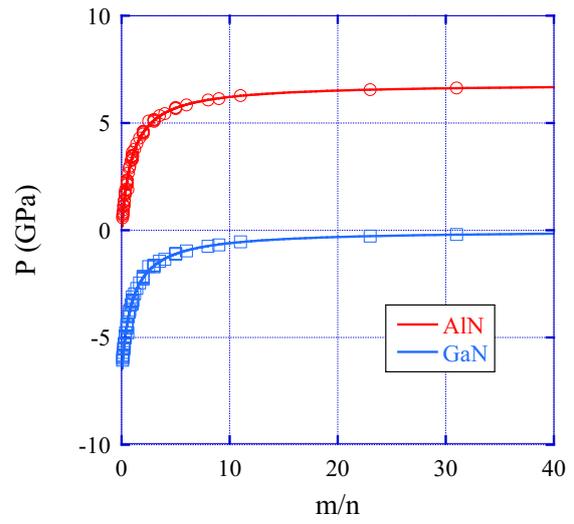

(a) (b)

**Figure 5** – Tangential stress components (a) and the corresponding hydrostatic pressures (b) as functions of m/n. Open circles and squares represent total energy calculations via Hooke's law, and solid lines the elastic prediction (see section III).

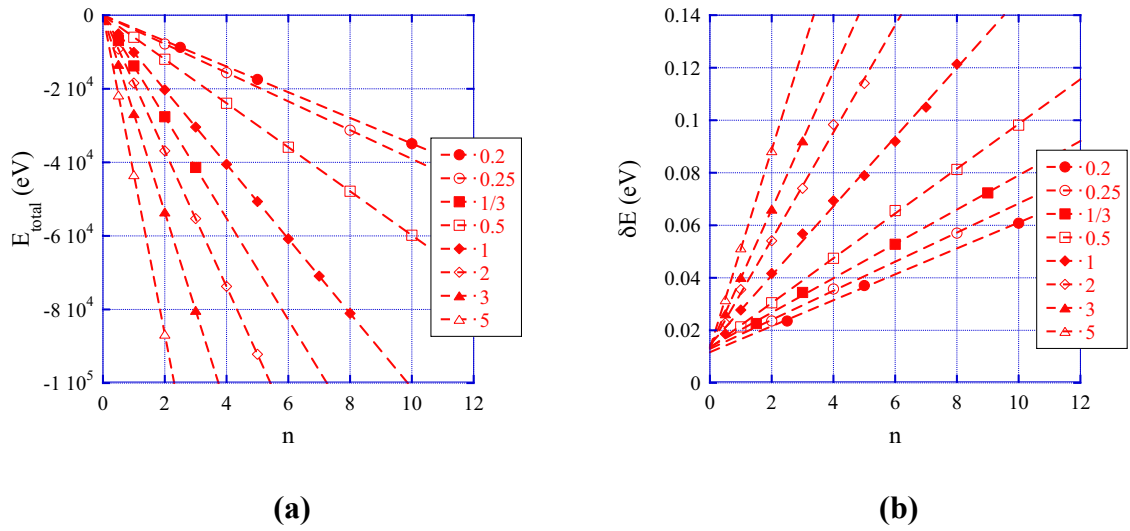

**Figure 6** – At a constant ratio of m/n, where m and n represent the thicknesses of GaN and AlN in terms of c-lattice cells, the total $E_{total}$ (equ. 11) and excess energies $\delta E$ of the superlattices (equ. 12) are linear functions of n. The investigated values of m/n are reported in the inserts. Dashed straight lines represent linear least-squares fits to the data.